# Gas modulation refractometry (GAMOR) – On its ability to eliminate the influence of drifts

OVE AXNER[1], ISAK SILANDER[1], CLAYTON FORSSÉN[1], YIN GUO[1], AND MARTIN ZELAN[2]

[1] Department of Physics, Umeå University, SE-901 87 Umeå, Sweden
[2] Measurement Science and Technology, RISE Research Institutes of Sweden, SE-501 15 Borås, Sweden

E-mail: ove.axner@umu.se

## Abstract

Gas modulation refractometry (GAMOR) is a technique based on a dual-Fabry-Perot (FP) cavity (DFPC) for assessment of gas refractivity, density, and pressure that can alleviate significant limitations of conventional refractometry systems, predominantly those related to drifts. Repeated assessments of the beat frequency when the measurement cavity is evacuated provide conditions under which the methodology is immune to the linear parts of the drifts in the system, both those from length changes of the cavities and those from gas leaks and outgassing. This implies that the technique is solely influenced by the non-linear parts of the drifts. This work provides a description of the principle behind the GAMOR methodology and explicates the background to its unique property. Based on simple models of the drifts of the temperature in the cavity spacer and the residual gas in the reference cavity, this work predicts that a GAMOR system, when used for assessment of refractivity, can sustain significant temperature drifts and leakage rates without being affected by noticeable errors or uncertainties. The cavity spacer can be exposed to temperature fluctuations of 100 mK over $10^3$ s, and the reference cavity can have a leakage that fills it up with gas on a timescale of days, without providing errors or uncertainties in the assessment of refractivity that are $3 \times 10^{-12}$, which, for $N_2$, corresponds to 0.01 ppm (parts per million) of the value under atmospheric pressure conditions, and thereby 1 mPa. Since well-designed systems often have temperature fluctuations and leakage rates that are smaller than these, it is concluded that there will, in practice, not be any appreciable influence from cavity length drifts, gas leaks, and outgassing in the GAMOR methodology.

## 1. Introduction

Refractometry constitutes a technique for assessment of refractivity [1-4]. When performed under controlled conditions, it has been shown that it can be used to assess gas density and pressure with high precision [1-16]. Recent works have also indicated that the technique has the potential to replace current pressure standards, in particular in the 1 to 100 kPa range [17-23].

A common means to realize refractometry is to base it on a Fabry-Perot (FP) cavity (FPC) [1, 2, 24-26]. When gas whose refractivity is to be assessed is let into such a cavity, the frequencies of its cavity modes will shift [2, 3]. By locking a laser (referred to as the measurement laser) to one of these modes, the shift of the frequency of the mode addressed will be transferred to a shift in the frequency of the laser light [2, 3].

A common way to assess such a shift is to mix the frequency of the laser light down to a RF frequency by use of another laser (a reference laser). This is achieved by merging the two laser fields onto a photodiode that can directly measure the beat frequency between them. By this, the shift in the frequency of the cavity mode addressed is converted to a shift in the measured beat frequency.

Although it is simple in theory to realize FP-based instrumentation for refractometry, it is not trivial in practice. One reason is that such cavities are influenced by various types of drifts, predominantly those from changes of the length of the cavity (which, in turn, mainly originate from alterations in the temperature of the cavity spacer or from material creeping), gas leaks, and outgassing [16, 27-30]. For example, a drift in length of 1 pm of a 30 cm long cavity during a measurement corresponds to an error in the assessment of refractivity and pressure of $3 \times 10^{-12}$ and (for $N_2$) 1 mPa, respectively. This implies that FP-based refractometers, in their basic realization, here referred to as single-FPC (SFPC)-based refractometry, require extremely good temperature control and assessment capabilities.

To alleviate some of these shortcomings, refractometry is often based on a dual-Fabry-Perot cavity (DFPC) in which two cavities, bored in the same spacer block (often referred to as the measurement and the reference cavity, respectively), are simultaneously addressed by two laser fields [2, 3, 5, 10, 14-17, 29, 30]. This has two major advantages.

One is that any change in length of the cavity spacer that is common to the two cavities will cancel in the measured beat frequency whereby it will not influence any assessment of refractivity. Another is that the reference laser achieves the same stability and frequency width as the measurement laser, which facilitates the assessment of the beat frequency.

Despite this, since two cavities in a given spacer can drift dissimilarly, DFPC-based refractometry is still influenced by cavity drifts [29]. A possible remedy to this is to utilize a detection methodology that is affected as little as possible of the drifts. As a means to achieve this, Gas Modulation Refractometry (GAMOR) has recently been developed [31, 32]. GAMOR is a methodology that is immune to large parts of the drifts in refractometry systems, not only those from changes in length of the cavity (e.g. those caused by drifts in the temperature of the spacer material) but also those from gas leaks or outgassing into the reference cavity [31]. In its first realization, it was demonstrated that the technique is capable of reducing drifts of the refractometry signal from a DFPC-based system with no active temperature control by up to three orders of magnitude, from the Pa to the mPa range [31].

This unique and important property originates from a measurement procedure in which the refractivity of the gas in the measurement cavity is assessed from the difference between the beat frequency measured with gas in the measurement cavity and an interpolation between two empty measurement cavity beat frequency assessments. By this, the linear parts of the drifts are eliminated, making a system based on the GAMOR methodology immune to the dominating parts of the drifts [31].[1]

However, if the drifts also have a non-linear behaviour, these will not, in the conventional realization of GAMOR, be picked up by the interpolation procedure, and thereby not be eliminated. This can potentially give rise to minor errors or uncertainties. It has not yet been estimated or assessed how large these can be from typical types of drifts and under which conditions they will influence a GAMOR assessment. This is an unsatisfactory situation that needs to be remedied. This work therefore provides an assessment of how non-linear drifts influence the GAMOR methodology and an estimate of their influence.

In order to do this, it first provides a basic description of the general principles of the GAMOR methodology. To illustrate its benefits, it starts by comparing the limiting processes of GAMOR with those of other types of FC-based refractometry, *viz.* (conventional) SFPC-based refractometry and conventional (unmodulated) DFPC-based refractometry. It is shown that while SFPC-based refractometry is limited by the drifts in the measurement cavity, conventional DFPC-based refractometry is restricted by the difference in drift of the two cavities. GAMOR, finally, is not affected by any of the linear parts of the drift in any of the cavities; it is solely affected by the difference in the non-linear parts of the drifts of the two cavities.

To assess to which degree drifts still can influence the GAMOR methodology, this work then estimates under which conditions the most common causes of drifts, i.e. those of the lengths of the cavities caused by temperature fluctuations and those due to gas leakage and outgassing (primarily in the reference cavity), will influence (i.e. provide errors or uncertainties in) assessments below either a given "strict" or a stated "relaxed" benchmark. These represent errors (or uncertainties) in refractivity that are $3 \times 10^{-12}$ and $3 \times 10^{-10}$, which, for $N_2$ correspond to $10^{-8}$ (or 0.01 ppm, parts per million) and $10^{-6}$ (1 ppm) of their values under atmospheric

---

[1] Note that the technique cannot eliminate the influence of deformations of the cavities due the presence of gas. Such phenomena are therefore not addressed in this work; they are dealt with elsewhere.

pressure conditions (thus representing pressures of 1 mPa and 0.1 Pa), respectively.

Based upon general models for the two dominating types drifts, *viz.* thermal induced drifts of the lengths of the cavities and gas leaks and outgassing, it is then shown that the cavity spacer material can experience temperature fluctuations that are in the order of 100 mK over $10^3$ s, or, the reference cavity can have a leak (or be exposed to outgassing) that fills it up with gas to $1-e^{-1}$ of the surrounding pressure on a timescale of 4 days, before the instrumentation gives rise to errors that correspond to the strict benchmark. The corresponding numbers for the relaxed benchmark are one order of magnitude larger and smaller, *viz.* 1 K and 10 hours, respectively.

Since well-designed systems often have temperature fluctuations and leakage (or outgassing) rates far below these, it is concluded that, when such a GAMOR system is used, there will, in practice, not be any appreciable influence of drifts to refractivity assessments.

## 2. Theory

The basis for GAMOR is the same as that of both SFPC-based refractometry and conventional DFPC-based refractometry. In all these types of technique, the refractivity of a gas in an FP cavity is assessed in terms of the change in frequency of a mode of the cavity as the cavity is evacuated. Their differences constitute the way in which the reference and the measurement assessments are carried out. The defining expressions for FP-based refractometry, which are common to all three types of technique, are therefore first derived. To place the properties of GAMOR in perspective of those of the other types of realization of FP-based refractivity, the properties of all these types of technique are therefore first compared.

### 2.1. Nomenclature and expressions for assessment of refractivity by FP-based refractometry

#### 2.1.1. Nomenclature.

For all types of technique, we will assume that the gas whose refractivity is to be assessed (assumed to be $n_g$) is let into an FP cavity, henceforth referred to as the measurement cavity, denoted cavity "$m$", whose length, when being empty, is $L_m^{(0)}$. A laser (referred to as the measurement laser) is then addressing (in reality, locked to) the $q_m^{(0)}$ :th longitudinal mode of the cavity. By this, the frequency of the laser light becomes equal to that of the cavity mode addressed, $\nu_m^{(0)}$, which is given by

$$\nu_m^{(0)} = \frac{q_m^{(0)} c}{2 L_m^{(0)}}. \tag{1}$$

When gas is let into such a cavity, the frequency of the laser light will change due to three reasons: *viz.* an increase in refractivity of the gas in the cavity, from 0 to $n_g - 1$; an change in the length of the cavity (due to the pressure the gas exerts on the cavity), from $L_m^{(0)}$ to $L_m^{(0)} + \delta L_m^{(0 \to g)}$; and, if the laser performs mode hops during the gas filling process, an increase in the number of the mode addressed, from $q_m^{(0)}$ to $q_m^{(0)} + \Delta q_m^{(0 \to g)}$. As has been shown in the literature [3, 31, 32], this implies that the frequency of the laser light, when addressing the filled measurement cavity, $\nu_m^{(g)}$, can be expressed as

$$\nu_m^{(g)} = \frac{\left[q_m^{(0)} + \Delta q_m^{(0 \to g)}\right] c}{2 n_g \left[L_m^{(0)} + \delta L_m^{(0 \to g)}\right]} = \nu_m^{(0)} \frac{1 + \overline{\Delta q_m}^{(0 \to g)}}{n_g \left[1 + \overline{\delta L_m}^{(0 \to g)}\right]}, \tag{2}$$

where, in the last step, $\overline{\Delta q_m}^{(0 \to g)}$ and $\overline{\delta L_m}^{(0 \to g)}$ represent the corresponding relative entities, given by $\Delta q_m^{(0 \to g)} / q_m^{(0)}$ and $\delta L_m^{(0 \to g)} / L_m^{(0)}$, respectively.

When an assessment of reactivity is performed, it is customary to assess the difference between the laser frequencies when the measurement cavity is evacuated and filled with gas. In practice, however, the measurement cavity is not evacuated to full vacuum when the empty-cavity measurement is performed; instead, it contains a residual amount of gas, with a refractivity of $n_{Res} - 1$. The laser will then have a frequency $\nu_m^{(Res)}$ that can be expressed as

$$\nu_m^{(Res)} = \frac{\left[q_m^{(0)} + \Delta q_m^{(0 \to Res)}\right] c}{2 n_{Res} \left[L_m^{(0)} + \delta L_m^{(0 \to Res)}\right]} = \nu_m^{(0)} \frac{1 + \overline{\Delta q_m}^{(0 \to Res)}}{n_{Res} \left[1 + \overline{\delta L_m}^{(0 \to Res)}\right]}, \tag{3}$$

where the various entities have definitions that correspond to those for $\nu_m^{(g)}$ in Eq. (2).

As was alluded to above, a common means to assess a shift of the frequency of laser light is to mix it with that from another laser (a reference laser), down to a RF frequency that can be assessed by a fast photodiode. In SFPC-based refractometry, the reference laser is often a separate, well-stabilized laser (possibly a frequency comb) [11].

In both conventional DFPC-based refractometry and GAMOR, the reference laser addresses a mode of a second cavity in the cavity spacer (here for simplicity assumed to be empty) whose frequency, $\nu_r^{(0)}$, is given by an expression similar to that for the empty measurement cavity, given by

$$\nu_r^{(0)} = \frac{q_r^{(0)} c}{2 L_r^{(0)}}, \tag{4}$$

where the various entities have definitions that correspond to those for $\nu_m^{(0)}$ in Eq. (1).

Regardless of whether the reference laser is locked to a second cavity or not, we will henceforth denote the frequency of the reference laser $\nu_r^{(0)}$.

This implies that, for all methodologies considered, the beat frequencies measured when the measurement cavity is evacuated (i.e., when it contains a residual amount of gas) and when it is filled with gas, referred to as $f_{(0,Res)}$ and $f_{(0,g)}$, can be written as

$$f_{(0,Res)} \equiv v_r^{(0)} - v_m^{(Res)} = v_r^{(0)} - v_m^{(0)} \frac{1+\overline{\Delta q_m}^{(0 \rightarrow Res)}}{n_{Res}\left[1+\overline{\delta L_m}^{(0 \rightarrow Res)}\right]} \quad (5)$$

and

$$f_{(0,g)} \equiv v_r^{(0)} - v_m^{(g)} = v_r^{(0)} - v_m^{(0)} \frac{1+\overline{\Delta q_m}^{(0 \rightarrow g)}}{n_g\left[1+\overline{\delta L_m}^{(0 \rightarrow g)}\right]}, \quad (6)$$

respectively, where the first subscript, 0, indicates that the reference cavity is empty and the second one denotes the conditions for the measurement cavity.

*2.1.2. Expressions for the assessment of refractivity by FP-based refractometry*

As has been shown previously in the literature [3, 14, 31], for the case when $(n_{Res}-1) \ll (n_g-1)$, which normally is valid when refractometry techniques are used, from which it follows that $\overline{\delta L_m}^{(0 \rightarrow Res)} \ll \overline{\delta L_m}^{(0 \rightarrow g)}$, it is possible to express the refractivity of the gas in the measurement cavity, $n_g - 1$, as

$$n_g - 1 \approx \frac{\overline{\Delta f}_{(0,Res \rightarrow g)} + \overline{\Delta q_m}^{(Res \rightarrow g)}}{1 - \overline{\Delta f}_{(0,Res \rightarrow g)} + \varepsilon_m} + (n_{Res}-1), \quad (7)$$

where $\overline{\Delta f}_{(0,Res \rightarrow g)} = \Delta f_{(0,Res \rightarrow g)}/v_m^{(0)}$ represents the relative change in beat frequency when the measurement cavity is filled (or evacuated), where, in turn, $\Delta f_{(0,Res \rightarrow g)}$ is the shift in beat frequency, given by

$$\Delta f_{(0,Res \rightarrow g)} = f_{(0,g)} - f_{(0,Res)}. \quad (8)$$

Moreover, $\overline{\Delta q_m}^{(Res \rightarrow g)}$ denotes the number of mode jumps the laser performs when the gas is filled into the cavity (or when it is evacuated), given by $\overline{\Delta q_m}^{(0 \rightarrow g)} - \overline{\Delta q_m}^{(0 \rightarrow Res)}$, while $\varepsilon_m$ represents a refractivity-normalized deformation coefficient of the measurement cavity, defined as $\varepsilon_m(n_{Ext}-n_{Res}) = \delta L_m^{(Res \rightarrow Ext)}/L_m^{(0)}$, where, in turn, $\delta L_m^{(Res \rightarrow Ext)}$ is given by $\delta L_m^{(0 \rightarrow g)} - \delta L_m^{(0 \rightarrow Res)}$ [3].

Equation (7) shows that an assessment of $n_g - 1$ requires knowledge about the residual refractivity in the measurement cavity when it is evacuated, $n_{Res} - 1$. Since the residual amount of gas in this cavity is significantly smaller than that when it is filled with gas, $n_{Res} - 1$ provides only a minor contribution to $n_g - 1$. This implies that it does not have to be determined with as high relative accuracy as $n_g - 1$. It therefore suffices to assess $n_{Res} - 1$ by a measurement of the pressure in the cavity by a standard pressure gauge. This can then be used to recalculate the evacuated measurement cavity beat frequency, $f_{(0,Res)}$, to its corresponding empty measurement cavity beat frequency [3, 15, 31], $f_{(0,0)}$, which represents the beat frequency that would have been measured if the measurement cavity would have been evacuated to pure vacuum, defined as

$$f_{(0,0)} \equiv v_r^{(0)} - v_m^{(0)}, \quad (9)$$

by the use of

$$f_{(0,0)} = f_{(0,Res)} + (n_{Res}-1)v_m^{(0)}. \quad (10)$$

By this, Eq. (7) can be simplified to

$$n_g - 1 = \frac{\overline{\Delta f}_{(0,0 \rightarrow g)} + \overline{\Delta q_m}^{(Res \rightarrow g)}}{1 - \overline{\Delta f}_{(0,0 \rightarrow g)} + \varepsilon_m}, \quad (11)$$

where $\overline{\Delta f}_{(0,0 \rightarrow g)}$ is given by $\Delta f_{(0,0 \rightarrow g)}/v_m^{(0)}$, where, in turn, $\Delta f_{(0,0 \rightarrow g)}$ is given by

$$\Delta f_{(0,0 \rightarrow g)} = f_{(0,g)} - f_{(0,0)}. \quad (12)$$

Although both the Eqs. (7) and (11) can be used for SFPC-based and conventional DFPC-based refractometry, Eq. (11) is the one that in practice is used when GAMOR is performed [31, 32]. We will therefore base the analysis given below on this.

*2.2. Comparison of the types of drift three common types of refractometry are affected by*

*2.2.1. Means of how to address the expression for the refractivity of gas by the three types of refractometry*

Equation (11) is valid for assessment of refractivity in all three types of refractometry addressed in this introductory comparison (SFPC-based refractometry, conventional DFPC-based refractometry, and GAMOR). However, because of practical reasons, none of them will, in practice, be able to assess the refractivity of the gas in the measurement cavity without being affected by drifts in the system to some extent. The reason is that the shift in beat frequency, $\Delta f_{(0,0 \rightarrow g)}$, will not solely originate from the change in the refractivity of the gas in the measurement cavity, as indicated by Eq. (12); it will also be influenced by the drifts in the system that take place between the instances when the measurement cavity is filled with gas (i.e. when $f_{(0,g)}$ is assessed) and when it is evacuated (i.e. when $f_{(0,0)}$ is assessed).

Such drifts comprise predominantly alterations in the length of the cavities (originating from drifts of the temperature of the spacer material) [16, 27-30]. For the DFPC-based techniques, they can also be caused by gas leaks or outgassing into the reference cavity, which, to reduce the risk for unintentional fluctuations in its gas density, often is evacuated solely once in the beginning of a measurement series. Hence, neither $f_{(0,0)}$, nor $f_{(0,g)}$, will be constant over time, neither through an entire measurement series, nor during a single measurement cycle; they will "always" drift. It is therefore of importance to develop instrumentation and methodologies that

(i) ascertain that the two beat frequencies, $f_{(0,0)}$ and $f_{(0,g)}$, are affected as little as possible by the aforementioned drifts, and
(ii) assess them within an as short period of time as possible, ideally at the same time.

The major differences between the three types of refractometry addressed in this introductory comparison comprise how these two objectives are addressed.

As a mean to address the first constrain, DFPC-based refractometry was developed from SFPC-based refractometry. By addressing a second cavity in the cavity spacer material, drifts in the cavity spacer that affects the lengths of the two cavities equally, which are assumed to affect $v_r^{(0)}$ and $v_m^{(0)}$ similarly, will, to a large extent, cancel in both $f_{(0,0)}$ and $f_{(0,g)}$, and thereby also in $\Delta f_{(0,0 \to g)}$.

Regarding the latter constrain, since it is not technically possible to assess $f_{(0,0)}$ and $f_{(0,g)}$ simultaneously, in conventional DFPC refractometry these two entities are assessed at dissimilar time instances. To alleviate this, GAMOR was developed. In this methodology, a value of $f_{(0,0)}$ for the time $f_{(0,g)}$ was measured (denoted $t_g$) is estimated by an interpolation between two assessments of $f_{(0,0)}$, performed before and after the filled measurement cavity assessment. The value obtained, denoted $\tilde{f}_{(0,0)}(t_g)$, is considered to be an adequate representation of the $f_{(0,0)}$ value that represents the empty measurement cavity beat frequency that would have been at $t_g$ in case the measurement cavity had not been filled with gas.

In order to quantitatively assess to which extent the three different methodologies are affected by drifts, so as to properly assess the advantages and limitations of GAMOR, we will commence by defining a model for the drifts of the empty cavity mode frequencies: for SFPC refractometry, for the evacuated measurement cavity mode frequency, $v_m^{(0)}$; and, for the two DFPC-based refractometry techniques, for both $v_m^{(0)}$ and $v_r^{(0)}$.

*2.2.2. A model for the drifts of the cavity modes*

Let us assume that the empty cavity mode frequencies, for each cavity (*i = m* or *r*, representing the measurement and the reference cavity, respectively) and for each measurement cycle, can be written as Taylor series expanded around the time when the assessment of $f_{(0,g)}$ is made, i.e. $t_g$, as

$$v_i^{(0)}(t) = v_i^{(0)}(t_g) + \left(\frac{\partial v_i^{(0)}}{\partial t}\right)_{t_g}(t-t_g) \\ + \frac{1}{2}\left(\frac{\partial^2 v_i^{(0)}}{\partial t^2}\right)_{t_g}(t-t_g)^2 + O[(t-t_g)^3], \quad (13)$$

where $(\partial v_i^{(0)}/\partial t)_{t_g}$ and $(\partial^2 v_i^{(0)}/\partial t^2)_{t_g}$ are the first and second order derivatives of the frequency of the cavity mode addressed in cavity *i* at $t_g$.

This implies that the empty cavity measurement beat frequency, given by Eq. (9), will be time dependent, $f_{(0,0)}(t)$, and can be expressed as

$$f_{(0,0)}(t) = f_{(0,0)}(t_g) + \left(\frac{\partial f_{(0,0)}}{\partial t}\right)_{t_g}(t-t_g) \\ + \frac{1}{2}\left(\frac{\partial^2 f_{(0,0)}}{\partial t^2}\right)_{t_g}(t-t_g)^2 + O[(t-t_g)^3] \quad (14)$$

where $f_{(0,0)}(t_g) = v_r^{(0)}(t_g) - v_m^{(0)}(t_g)$, $(\partial f_{(0,0)}/\partial t)_{t_g} = (\partial v_r^{(0)}/\partial t)_{t_g} - (\partial v_m^{(0)}/\partial t)_{t_g}$, and $(\partial^2 f_{(0,0)}/\partial t^2)_{t_g} = (\partial^2 v_r^{(0)}/\partial t^2)_{t_g} - (\partial v_m^{(0)}/\partial t)_{t_g}$.

An example of a typical empty cavity measurement beat frequency is given by the blue solid curve in Figure 1.

*2.2.3. The error in the assessment of the beat frequency from drifts for SFPC-based refractometry*

In SFPC based refractometry, the shift in the beat frequency is given by Eq. (12), which, when the time instances for the assessments are explicitly stated, reads

$$\Delta f_{(0,0 \to g)}(t_g, t) = f_{(0,g)}(t_g) - f_{(0,0)}(t), \quad (15)$$

where thus $f_{(0,g)}(t_g)$ is the beat frequency when the measurement cavity contains the gas whose refractivity is to be assessed at the time $t_g$ while $f_{(0,0)}(t)$ is the empty cavity measurement beat frequency assessed at a different time *t*. Due to drifts, this shift differs from the one that preferably should be used in Eq. (11) for the most accurate assessment, *viz.* the one when $f_{(0,g)}$ and $f_{(0,0)}$ are assessed at the same time, denoted $\Delta f_{(0,0 \to g)}(t_g, t_g)$ and defined as

$$\Delta f_{(0,0 \to g)}(t_g, t_g) = f_{(0,g)}(t_g) - f_{(0,0)}(t_g). \quad (16)$$

This implies that the error in the assessment of the beat frequency, defined as the difference between the latter two entities, i.e. $\Delta f_{(0,0 \to g)}(t_g, t_g) - \Delta f_{(0,0 \to g)}(t_g, t)$, denoted $\delta[\Delta f(t)]$, is, for SFPC based refractometry, given by

$$\delta[\Delta f(t)] \equiv \Delta f_{(0,0 \to g)}(t_g, t_g) - \Delta f_{(0,0 \to g)}(t_g, t) \\ = f_{(0,0)}(t) - f_{(0,0)}(t_g), \quad (17)$$

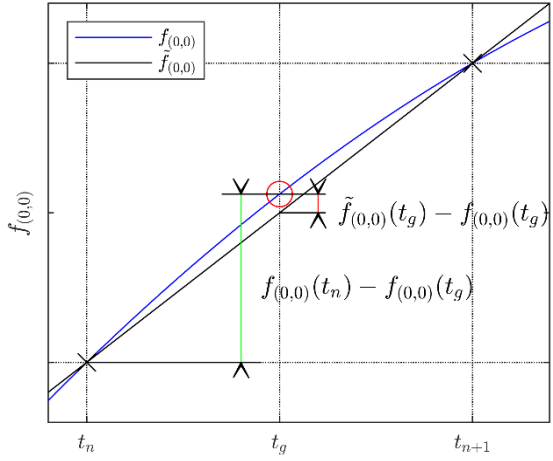

**Figure 1.** Blue solid curve: the empty measurement cavity beat frequency in the presence of drifts, $f_{(0,0)}(t)$, as given by Eq. (14). The beat frequency at time $t_g$, $f_{(0,0)}(t_g)$, is marked by a red circle. Black straight line: estimated empty measurement cavity beat frequency, $\tilde{f}_{(0,0)}(t)$, created by a linear interpolation between two empty measurement cavity beat frequency assessments (marked by crosses), according to an expression similar to that given in Eq. (22). The small gap, represented by the red line, denoted $\tilde{f}_{(0,0)}(t_g) - f_{(0,0)}(t_g)$, marks the error in the assessment of the beat frequency, $\delta[\Delta f(t_n, t_g, t_{n+1})]$, given by Eq. (25), while the large gap, marked by a green line, denoted $f_{(0,0)}(t_n) - f_{(0,0)}(t_g)$, illustrates the error made when conventional DFPC-based refractometry is performed, as given by Eq. (21). Note that, in this example, both errors take negative values. The crosses mark the positions of the reference assessments, $[t_n, f_{(0,0)}(t_n)]$ and $[t_{n+1}, f_{(0,0)}(t_{n+1})]$, respectively. The curvature of $f_{(0,0)}(t)$ is, for most practical cases, greatly exaggerated for illustrative purposes.

where $f_{(0,0)}(t_g)$ and $f_{(0,0)}(t)$ thus represent the actual and the measured empty measurement cavity beat frequencies at the times $t_g$ and $t$, respectively.

Since the reference laser in SFPC-based refractometry is not assessing any cavity mode, its frequency will not be affected by any drift influencing the cavity modes. This implies that Eq. (17) can be written as

$$\delta[\Delta f(t)] = \left[ v_r^{(0)} - v_m^{(0)}(t) \right] - \left[ v_r^{(0)} - v_m^{(0)}(t_g) \right] \\ = v_m^{(0)}(t_g) - v_m^{(0)}(t). \quad (18)$$

Inserting Eq. (13) into this implies that $\delta[\Delta f(t)]$ can be written as

$$\delta[\Delta f(t)] = -\left( \frac{\partial v_m^{(0)}}{\partial t} \right)_{t_g} (t - t_g) + O\left[(t-t_g)^2\right]. \quad (19)$$

This shows that the error in SFPC-based refractometry is given by a product of the first order derivative of the frequency of the empty measurement cavity (the linear drift), $(\partial v_m^{(0)} / \partial t)_{t_g}$, and the separation in time between the assessments of $f_{(0,0)}$, and $f_{(0,g)}$, $(t-t_g)$.

### 2.2.4. The error in the assessment of the beat frequency from drifts for conventional DFPC-based refractometry

In conventional DFPC-based refractometry, the error in the assessment of the beat frequency is likewise given by Eq. (17). In this case though, the reference laser is addressing the reference cavity of the spacer material. This implies that the error in the assessment of the beat frequency becomes

$$\delta[\Delta f(t)] = \left[ v_r^{(0)}(t) - v_m^{(0)}(t) \right] - \left[ v_r^{(0)}(t_g) - v_m^{(0)}(t_g) \right] \\ = \left[ v_m^{(0)}(t_g) - v_m^{(0)}(t) \right] - \left[ v_r^{(0)}(t_g) - v_r^{(0)}(t) \right]. \quad (20)$$

Inserting Eq. (13) into this implies that the error in the assessment of the beat frequency can be written as

$$\delta[\Delta f(t)] = -\left[ \left( \frac{\partial v_m^{(0)}}{\partial t} \right)_{t_g} - \left( \frac{\partial v_r^{(0)}}{\partial t} \right)_{t_g} \right](t - t_g) \\ + O\left[(t-t_g)^2\right]. \quad (21)$$

This shows that in conventional DFPC-based refractometry the error is given by a product of the difference in the first order derivatives of the frequency of the empty cavities, $(\partial v_m^{(0)} / \partial t)_{t_g} - (\partial v_r^{(0)} / \partial t)_{t_g}$, and the separation in time between the $f_{(0,0)}$ and $f_{(0,g)}$ assessments, $(t - t_g)$.

A comparison with Eq. (19) reveals that the advantage of DFPC-based refractometry is that in DFPC-based refractometry the error is only proportional to the *difference* in drift of the frequencies of the two empty cavities while in SFPC-based refractometry it is proportional to the entire drift of the empty measurement cavity.

This shows that if a cavity spacer can be made so that the differential drift is significantly smaller than the common drift, DFPC-based refractometry will be significantly less affected by drifts than SFPC-based refractometry.

### 2.2.5. The error in the assessment of the beat frequency from drifts for GAMOR

However, conventional DFPC-based refractometry is still limited by the fact that $f_{(0,0)}$ cannot be assessed at the same time as $f_{(0,g)}$. GAMOR was developed to alleviate this. Figure 2 shows an illustration of the principle of GAMOR. As is shown in Figure 2(a), the measurement cavity is repeatedly evacuated and filled with gas. As then is displayed in Figure 2(c), the empty measurement cavity beat frequency (the green line) is estimated at all times during such a cycle by the use of a linear interpolation between two empty measurement cavity beat frequency assessments — one performed directly prior to when the measurement cavity is filled with gas during each measurement cycle, for cycle *n*, at a time denoted $t_n$, and the other directly after the cavity has been evacuated, which is identical to the one carried out before the filling of cycle *n+1*, hence at a time denoted $t_{n+1}$, (both marked by crosses). This implies that the value of the empty measurement cavity beat frequency at the time $t_g$, denoted $\tilde{f}_{(0,0)}(t_n,t_g,t_{n+1})$, can be estimated, for cycle *n*, for which $t_n \leq t_g \leq t_{n+1}$, as

$$\tilde{f}_{(0,0)}(t_n,t_g,t_{n+1}) = f_{(0,0)}(t_n) \\ + \frac{f_{(0,0)}(t_{n+1}) - f_{(0,0)}(t_n)}{t_{n+1} - t_n}(t_g - t_n), \quad (22)$$

marked by a green circle in Figure 2(c).

This implies that the error in the assessment of the beat frequency in GAMOR, denoted $\delta[\Delta f(t_n,t_g,t_{n+1})]$, is given by

$$\delta[\Delta f(t_n,t_g,t_{n+1})] = \tilde{f}_{(0,0)}(t_n,t_g,t_{n+1}) - f_{(0,0)}(t_g). \quad (23)$$

Inserting Eq. (22) into this implies that $\delta[\Delta f(t_n,t_g,t_{n+1})]$ can be written as

$$\delta[\Delta f(t_n,t_g,t_{n+1})] = f_{(0,0)}(t_n) \\ + \frac{f_{(0,0)}(t_{n+1}) - f_{(0,0)}(t_n)}{t_{n+1} - t_n}(t_g - t_n) \quad (24) \\ - f_{(0,0)}(t_g),$$

By use of the Eqs. (9) and (13), this expressions can be written, after some algebra, as

$$\delta[\Delta f(t_n,t_g,t_{n+1})] \approx -\frac{1}{2}\left[\left(\frac{\partial^2 v_m^{(0)}}{\partial t^2}\right)_{t_g} - \left(\frac{\partial^2 v_r^{(0)}}{\partial t^2}\right)_{t_g}\right] \quad (25) \\ \times (t_g - t_n)(t_{n+1} - t_g).$$

This shows, first of all, the important fact that the linear terms in the expressions for the mode frequencies in Eq. (13), which comprise the leading term in the expression for the error in the assessment of the beat frequency for conventional DFPC-based refractometry, given by Eq. (21), cancel, and do not contribute to the error of the GAMOR signal. This is the basis for GAMOR; the methodology eliminates the influence of the linear drifts of the cavity modes, making the technique immune to the dominating linear parts of the drifts. Moreover, Eq. (25) reveals that it is only affected by the *difference* in non-linearities in the drifts of the two cavity modes. This is the main reason why GAMOR has demonstrated three-order-of-magnitude smaller drifts than conventional DFPC-based refractometry when applied to a non-temperature-stabilized cavity spacer [31].

### 2.3 The influence of drifts on GAMOR

#### 2.3.1. Evaluation of the non-linear parts of the drifts of the cavity modes that affect the GAMOR signal

To assess the influence of the non-linear parts of the drifts of the cavity mode frequencies on GAMOR, it is necessary to allow for drifts both in the physical lengths of the cavities and their residual refractivities. A convenient means to model this is to assume that the reference cavity is not completely evacuated, as indicated by Eq. (4), but rather contains gas with a refractive index that slowly changes over time, denoted $n_r^{(Res)}(t)$. To allow for drifts of the length of the same cavity implies that we will simply express $L_r^{(0)}$ as $L_r^{(0)}(t)$. Moreover, to facilitate the derivation, we will denote the optical length of the cavity, $n_r^{(Res)}(t)L_r^{(0)}(t)$, by $\Lambda_r(t)$. This implies that the frequency of the reference cavity, $v_r^{(0)}(t)$, can be expressed as

$$\nu_r^{(0)}(t) = \frac{q_r^{(0)} c}{2\Lambda_r(t)}. \quad (26)$$

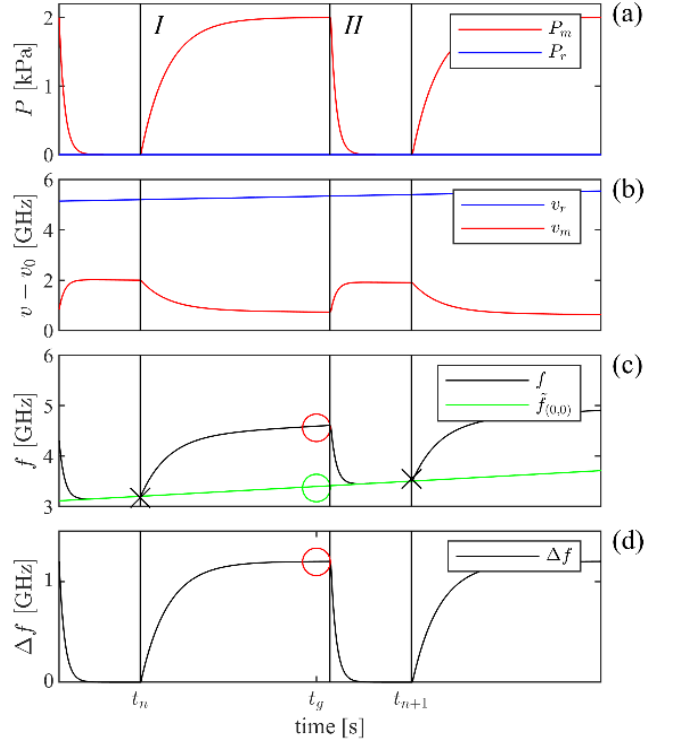

**Figure 2.** An illustration of the GAMOR principle displayed over two full measurement cycles. Panel (a) displays, as functions of time, the pressures in the measurement cavity, $P_m(t)$, (red curve) and in the reference cavity, $P_r(t)$, (blue curve). Panel (b) shows the corresponding frequencies of the measurement and reference lasers, $v_m(t)$ (red curve) and $v_r(t)$ (blue curve), respectively, (for display purposes, in the absence of mode jumps and offset to a common frequency $v_0$). Panel (c) illustrates the corresponding beat frequencies: the measured one, $f(t)$, (black curve), and the estimated empty measurement cavity beat frequency, $\tilde{f}_{(0,0)}(t_n,t,t_{n+1})$, (green line), according to an expression similar to Eq. (22). The slope of the latter is greatly exaggerated for display purposes. The red circle illustrates $f_{(0,g)}(t_g)$ while the green one indicates $\tilde{f}_{(0,0)}(t_g)$. The empty cavity beat frequencies are measured at the crosses. Panel (d), finally, displays the cavity-drift-corrected shift in beat frequency, $\Delta f(t)$, (black curve), at each time instance given by the difference between $f(t)$ and the interpolated $\tilde{f}_{(0,0)}(t_n,t,t_{n+1})$. The red circle indicates when the filled measurement cavity beat frequency typically is evaluated.

We will utilize a similar description for the evacuated measurement cavity, i.e. write it as

$$\nu_m^{(0)}(t) = \frac{q_m^{(0)} c}{2\Lambda_m(t)} \frac{1+\overline{\Delta q_m^{(0 \to Res)}}}{\left[1+\overline{\delta L_m^{(0 \to Res)}}\right]}, \quad (27)$$

where we have used $\Lambda_m(t)$ to denote $n_m^{(Res)}(t) L_m^{(0)}(t)$.

According to Eq. (25), to assess the error in the assessment of the beat frequency in GAMOR, the second derivatives of these entities are needed. Based on the expressions above, their first and second order derivatives can be evaluated as

$$\left(\frac{\partial \nu_i^{(0)}(t)}{\partial t}\right)_{t_g} = \left(\frac{\partial \nu_i^{(0)}}{\partial \Lambda_i}\right)_{t_g} \left(\frac{\partial \Lambda_i(t)}{\partial t}\right)_{t_g}$$
$$= -\left(\frac{q_i c}{2\Lambda_i^2}\right)_{t_g} (\dot{\Lambda}_i)_{t_g} = -\nu_i^{(0)}(t_g)\left(\frac{\dot{\Lambda}_i}{\Lambda_i}\right)_{t_g} \quad (28)$$

and

$$\left(\frac{\partial^2 \nu_i^{(0)}(t)}{\partial t^2}\right)_{t_g} = \nu_i^{(0)}(t_g)\left(\frac{2\dot{\Lambda}_i^2 - \ddot{\Lambda}_i \Lambda_i}{\Lambda_i^2}\right)_{t_g}, \quad (29)$$

respectively, where we have denoted the first and the second time derivatives of $\Lambda_i$ by $\dot{\Lambda}_i$ and $\ddot{\Lambda}_i$, respectively. To facilitate the analysis, we have, since we deal with errors, approximated $[1+\overline{\Delta q_m^{(0 \to Res)}}]/[1+\overline{\delta L_m^{(0 \to Res)}}]$ by unity and denoted $\nu_i^{(0)}(t_g)$ by $\nu_i^{(0)}$ (since it in practice can be evaluated at any time instance in a measurement cycle).

Moreover, utilizing the fact that $\dot{\Lambda}_i$ and $\ddot{\Lambda}_i$ can be written as

$$\dot{\Lambda}_i = \dot{n}_i^{(Res)} L_i^{(0)} + n_i^{(Res)} \dot{L}_i^{(0)} \quad (30)$$

and

$$\ddot{\Lambda}_i = \ddot{n}_i^{(Res)} L_i^{(0)} + 2\dot{n}_i^{(Res)} \dot{L}_i^{(0)} + n_i^{(Res)} \ddot{L}_i^{(0)} \quad (31)$$

implies that the second order derivatives of the cavity mode frequencies for the two evacuated cavities can be written as

$$\left(\frac{\partial^2 \nu_i^{(0)}}{\partial t^2}\right)_{t_g} = \nu_i^{(0)} \left\{ \frac{2\left[\dot{n}_i^{(Res)}\right]^2 - \ddot{n}_i^{(Res)} n_i^{(Res)}}{\left[n_i^{(Res)}\right]^2} \right.$$
$$\left. +2\frac{\dot{n}_i^{(Res)}}{n_i^{(Res)}}\frac{\dot{L}_i^{(0)}}{L_i^{(0)}} + \frac{2(\dot{L}_i^{(0)})^2 - \ddot{L}_i^{(0)} L_i^{(0)}}{(L_i^{(0)})^2} \right\}_{t_g}. \quad (32)$$

### 2.3.2. The influence of the non-linear parts of the drifts of the cavity modes on the GAMOR signal

Since leakage and outgassing rates often are assessed in terms of pressure, it is convenient to rewrite the refractivity in terms of the pressure, $P$, by use of the linearized Lorentz-Lorenz equation and the ideal gas law, i.e. $\rho = (2/3A_R)(n-1)$ and $P = k_b T N_A \rho$, where $A_R$, $k_b$, $T$, and $N_A$, are the molar polarizability, the Boltzmann's constant, the temperature, and Avogadro's number, respectively, as

$$P_i^{(Res)} = \frac{2k_b T N_A}{3A_R}(n_i^{(Res)} - 1) = \beta(n_i^{(Res)} - 1), \quad (33)$$

where we have neglected possible virial coefficients and, in the second step, for simplicity, denoted the combination of entities in front of $(n_i^{(Res)} - 1)$ by $\beta$. This implies that $\dot{n}_i^{(Res)}$ and $\ddot{n}_i^{(Res)}$ can be written as $\dot{P}_i^{(Res)}/\beta$ and $\ddot{P}_i^{(Res)}/\beta$, respectively.

Making use of these expressions, and the fact that we in an expression of an error can approximate $n_i^{(Res)}$ by unity, gives

$$\left(\frac{\partial^2 \nu_i^{(0)}}{\partial t^2}\right)_{t_g} = \nu_i^{(0)} \left[ 2\left(\frac{\dot{L}_i^{(0)}}{L_i^{(0)}}\right)^2 - \frac{\ddot{L}_i^{(0)}}{L_i^{(0)}} \right.$$
$$\left. +2\frac{\dot{L}_i^{(0)}}{L_i^{(0)}}\frac{\dot{P}_i^{(Res)}}{\beta} + 2\left(\frac{\dot{P}_i^{(Res)}}{\beta}\right)^2 - \frac{\ddot{P}_i^{(Res)}}{\beta} \right]_{t_g}. \quad (34)$$

This implies that the error in the assessment of the change of the beat frequency in the GAMOR methodology, $\delta[\Delta f(t_n, t_g, t_{n+1})]$, can be expressed as

$$\delta[\Delta f(t_n, t_g, t_{n+1})]$$
$$\approx -\frac{1}{2}\nu_m^{(0)} \left\{ 2\left[\left(\frac{\dot{L}_m^{(0)}}{L_m^{(0)}}\right)^2 - \left(\frac{\dot{L}_r^{(0)}}{L_r^{(0)}}\right)^2\right] - \left[\frac{\ddot{L}_m^{(0)}}{L_m^{(0)}} - \frac{\ddot{L}_r^{(0)}}{L_r^{(0)}}\right] \right.$$
$$+2\left(\frac{\dot{L}_m^{(0)}}{L_m^{(0)}}\frac{\dot{P}_m^{(Res)}}{\beta} - \frac{\dot{L}_r^{(0)}}{L_r^{(0)}}\frac{\dot{P}_r^{(Res)}}{\beta}\right) \quad (35)$$
$$+2\left[\left(\frac{\dot{P}_m^{(Res)}}{\beta}\right)^2 - \left(\frac{\dot{P}_r^{(Res)}}{\beta}\right)^2\right]$$
$$\left. -\left[\frac{\ddot{P}_m^{(Res)}}{\beta} - \frac{\ddot{P}_r^{(Res)}}{\beta}\right] \right\}_{t_g} (t_g - t_n)(t_{n+1} - t_g),$$

where we have considered $\nu_m^{(0)} = \nu_r^{(0)}$.

To estimate the influence of the non-linearities of the drifts of the lengths of the cavities and the residual pressure of the gas in the cavities we need to consider the processes that cause these drifts in some detail.

Based on simple models for the dominating causes for drifts of cavity modes, *viz.* drifts of the lengths of the cavities and gas leaks and outgassing, we will conclude under which conditions the GAMOR methodology will experience an error (or provide an uncertainty, see below) that is below a given benchmark. We will investigate both when the methodology will experience an error that is below a so called "strict" benchmark, which is taken as a refractivity of 3 x $10^{-12}$, which, for N$_2$, corresponds to a pressure of 1 mPa, both representing a fraction of 10$^{-8}$ (or 0.01 ppm) of their values under atmospheric pressure conditions, and a "relaxed" benchmark, which represents a refractivity of $3 \times 10^{-10}$, corresponding to a pressure of 0.1 Pa, both representing a fraction of 10$^{-6}$ (or 1 ppm) of their values under atmospheric pressure conditions.

## 3. Drifts of the cavity lengths

### 3.1. A model for the drifts of the lengths of the cavities

To assess the influence of drifts of the length of the cavities, let us first assume that the cavity lengths are predominantly influenced by the local temperature of the spacer material, i.e. for cavity *i*, the temperature around that cavity, $T_i(t)$, whereby its length can be written as

$$L_i^{(0)}(T) = L_{i,0}^{(0)} \left\{ 1 + \alpha \left[ T_i(t) - T_{i,0} \right] \right\}, \quad (36)$$

where $L_{i,0}^{(0)}$ is the length of cavity at the temperature $T_{i,0}$ and $\alpha$ is the thermal expansion coefficient for the cavity spacer material.[2]

In the ideal case, the regulation of the temperature strives for keeping the temperature around the cavities constant. However, due to disturbances from the surrounding, including those of the electronics in the regulation units, the temperature will fluctuate around the set temperature, both in time and space. Although such a drift can have a variety of forms, let us here assume that it can, for each cavity, be described in terms of two exponential functions, one that represents the disturbance that causes the temperature to drift away from the set temperature, of the type $\Delta T_i \left( 1 - e^{-t/\tau_T} \right)$, and one that corresponds to the temperature regulation process that steers the temperature back to its set temperature, modelled as $e^{-t/\tau_T}$. Let us for simplicity also assume that these two processes have the same characteristic time, $\tau_T$. This implies that we can represent the temperature around cavity *i* when the temperature is exposed to a disturbance as

$$T_i(t) = T_{i,0} + \Delta T_i \left( 1 - e^{-t/\tau_T} \right) e^{-t/\tau_T}$$
$$= T_{i,0} + \Delta T_i \left( e^{-t/\tau_T} - e^{-2t/\tau_T} \right). \quad (37)$$

This implies that the length of the cavities can be written as

$$L_i^{(0)}(t) = L_{i,0}^{(0)} \left[ 1 + \alpha \Delta T_i \left( e^{-t/\tau_T} - e^{-2t/\tau_T} \right) \right]. \quad (38)$$

Under these conditions, $(\dot{L}_i^{(0)} / L_i^{(0)})^2$ and $\ddot{L}_i^{(0)} / L_i^{(0)}$ become

$$\left( \frac{\dot{L}_i^{(0)}(t)}{L_i^{(0)}} \right)^2 = \left( \frac{\alpha \Delta T_i}{\tau_T} \right)^2 \left( -e^{-t/\tau_T} + 2e^{-2t/\tau_T} \right)^2 \quad (39)$$

and

$$\frac{\ddot{L}_i^{(0)}(t)}{L_i^{(0)}} = \frac{\alpha \Delta T_i}{\tau_T^2} \left( e^{-t/\tau_T} - 4e^{-2t/\tau_T} \right), \quad (40)$$

respectively.

### 3.2. Estimate of the conditions for drifts of the cavity lengths to provide errors and uncertainties that are equal to the benchmarks

As can be seen from Eq. (35), any drift in temperature that results in the same change in length of the two cavities will not influence an assessment of refractivity by GAMOR in any noticeable way. However, if the two cavities are affected by dissimilar temperature drifts, they might deform differently. In this case, the drifts will give rise to an error in the assessment.

Consider, as a worst case scenario, the situation when only one cavity is influenced by a temperature fluctuation. Since $\alpha \Delta T_i \ll 1$, the Eqs. (39) and (40) show that $(\dot{L}_i^{(0)} / L_i^{(0)})^2$ is negligible with respect to $(\ddot{L}_i^{(0)} / L_i^{(0)})$. This implies that the error in the assessment of the beat frequency from a temperature fluctuation in one of the cavities, denoted $\delta[\Delta f(t_n, t_g, t_{n+1})]$, can be expressed as

$$\delta[\Delta f(t_n, t_g, t_{n+1})] \approx -\frac{1}{2} v_m^{(0)} \frac{\alpha \Delta T}{\tau_T^2} \left( e^{-t/\tau_T} - 4e^{-2t/\tau_T} \right)$$
$$\times (t_g - t_n)(t_{n+1} - t_g). \quad (41)$$

A worst case scenario also comprises the situation when $(t_g - t_n)(t_{n+1} - t_g) = (t_{n+1} - t_n)^2 / 4$. Since $(t_g - t_n)$ normally is slightly longer than $(t_{n+1} - t_g)$, we will here, for simplicity approximate the product by $(t_{n+1} - t_n)^2 / 5$. Let us also, as a part of the worst case scenario, solely consider the case at the beginning of the disturbance [for which $t < \tau_T$, whereby $\exp(-t/\tau_L) \approx 1$]. Evaluating Eq. (41) under these conditions, for which $\delta[\Delta f(t_n, t_g, t_{n+1})]$ will be denoted $\delta[\Delta f(t_n, t_{n+1})]$, implies that the error in refractivity, denoted $\delta(n_m - 1)$, can be written as

$$\delta(n_m - 1) \approx \frac{\delta[\Delta f(t_n, t_{n+1})]}{v_m^{(0)}} \approx \frac{3\alpha \Delta T}{10 \tau_T^2} (t_{n+1} - t_n)^2. \quad (42)$$

For the case when the spacer material is made of Zerodur, for which $\alpha$ can be in the 10$^{-8}$ K$^{-1}$ range, and, for the case with a cycle time of 100 s, the condition for obtaining a

---

[2] It is also possible that the length of a cavity can be altered by sudden relaxations [29]. However, since such a process is significantly shorter than a measurement cycle, it will, in the GAMOR methodology, only provide a single outlier. Such data points can easily be rejected in the data evaluation process. This implies that GAMOR is virtually not influenced by also such processes.

refractivity assessment with an error below to the strict benchmark is that

$$\Delta T < 10^{-7} \tau_T^2. \qquad (43)$$

For the case when the disturbances in temperature have a characteristic time ($\tau_T$) of $10^3$ s, this implies that the conditions for the error to be below the strict benchmark is that $\Delta T$ should not exceed 100 mK. Since a well-stabilized cavity spacer in general has fluctuations over $10^3$ s that are below this, it can be concluded that drifts of the temperature on this time scale seldom will influence a GAMOR assessment of refractivity noticeably.

However, it can be noticed though that if the cavity would be exposed to faster disturbances, e.g. those with a characteristic time in the $10^2$ s range, the system will be more vulnerable to temperature drifts. Such a system, when assessing refractivity, can only tolerate drifts of the temperature up to around 1 mK if the error should be below the strict benchmark condition.

For the case with the relaxed benchmark condition, i.e. when the refractivity benchmark is solely taken as 1 ppm of the atmospheric conditions, the corresponding acceptable temperature fluctuations are, for the cases with characteristic times of $10^3$ and $10^2$ s, two orders of magnitude larger, i.e. 10 K and 100 mK, respectively.

The same conditions are valid when density is assessed.

A similar analysis for conventional DFPC-based refractometry, based on the Eqs. (21), (28), (30), and (38), shows that this technique, when used for assessment of refractivity or density, is associated by significantly more stringent requirements of the temperature stability; an assessment will experience an error below the strict benchmark when

$$\Delta T < 3 \times 10^{-6} \tau_T. \qquad (44)$$

This implies that, for conventional DFPC-based refractometry, and for the case with a characteristic time of $10^3$ s, a temperature stability of 3 mK, which is 30 times lower than for GAMOR, is needed.

Since a temporal fluctuation in temperature can be both positive and negative, and influence both the measurement and the reference cavity, and most measurement series last longer than the characteristic time, in a series of measurements, this type of effect will predominantly give rise to a distribution of errors that manifest themselves as an uncertainty in the assessed measurement value. Since the uncertainty obtained from such a series of measurements cycles, $U(n_m - 1)$, will be smaller than the maximum error from an individual measurement cycle, it is possible to conclude that the latter constitutes an upper limit for the uncertainty from a series of measurements, i.e.

$$U(n_m - 1) \leq \delta(n_m - 1) \qquad (45)$$

where $\delta(n_m - 1)$ is given by Eq. (42). This implies that the conditions for obtaining assessments with uncertainties below the benchmarks stated will be even more relaxed than those stated for assessments with maximum errors being below the benchmarks derived above.

When pressure is to be assessed, drifts of the temperature will influence the assessment also through the equation of state. As long as the drifts are picked up by the temperature probes, they will be corrected for by the use of this equation. However, since every assessment is associated by an uncertainty, the assessment can still be affected by the uncertainty of the assessments. An uncertainty in the assessment of gas temperature of $\delta(T_m)$ will give rise to a relative uncertainty in the assessment of pressure, i.e. $\delta(P_m)/P_m$, that is given by $\delta(T_m)/T_m$. This implies, for example, that for the case with an uncertainty in the assessment of gas temperature of 10 mK, the relaxed benchmark, for which $\delta(P_m)$ is 0.1 Pa, can be obtained only when pressures below 3 kPa are assessed. This illustrates that for high pressures (predominantly those in the kPa range, depending on conditions), the uncertainty in the assessment of pressure that originates from the uncertainty in the assessment of the temperature will dominate over those originating from drifts of the cavity lengths.

It is also possible that the length of a cavity can be altered by gas that diffuses into the spacer material. This is particularly the situation for He and ULE glass [33]. However, as for thermal expansion, since any such change in length is a slow process, considerable slower than the measurement cycle time, the GAMOR methodology will eliminate the influence of the linear parts of this elongation, whereby only the minor non-linear parts will remain.

## 4. Gas leaks and outgassing

### 4.1. A model for the drifts of the pressure in the reference cavity

To assess the influence of leaks (or outgassing) in the reference cavity, let us assume that the cavity has been evacuated in the beginning of a series of measurements, after which it has been sealed off, and that its pressure, $P_r^{(Res)}(t)$, thereafter increases monotonically with time as

$$P_r^{(Res)}(t) = P_{atm}\left(1 - e^{-t/\tau_r}\right), \qquad (46)$$

where $P_{atm}$ is the atmospheric pressure and $\tau_r$ is the characteristic time for the leakage into the reference cavity (representing the time to which the pressure in the cavity has increased to $1 - e^{-1}$ of that of the surrounding). We have here also tacitly assumed that the gas that leaks into the cavity has the same index of refraction as that already in the refractometer; the case when this does not hold is discussed further below.

In this case $\dot{P}_r^{(Res)}(t)$ and $\ddot{P}_r^{(Res)}(t)$ become

$$\dot{P}_r^{(Res)}(t) = \left(P_{atm}/\tau_r\right)e^{-t/\tau_r} \qquad (47)$$

and

$$\ddot{P}_r^{(Res)}(t) = -\left(P_{atm}/\tau_r^2\right)e^{-t/\tau_r}, \qquad (48)$$

respectively.

Since $\beta$ alternatively can be expressed as $P_{atm}/(n-1)_{atm}$, it is possible to express $[\dot{P}_r^{(Res)}(t)/\beta]^2$ and $\ddot{P}_r^{(0)}(t)/\beta$ in Eq. (35) as

$$\left(\frac{\dot{P}_r^{(Res)}(t)}{\beta}\right)^2 = \frac{(n-1)_{atm}^2}{\tau_r^2}e^{-2t/\tau_r} \qquad (49)$$

and

$$\frac{\ddot{P}_r^{(Res)}(t)}{\beta} = -\frac{(n-1)_{atm}}{\tau_r^2}e^{-t/\tau_r}, \qquad (50)$$

respectively.

## 4.2. The drifts of the residual pressure in the measurement cavity

Since the measurement cavity is not evacuated all the way to a complete vacuum between consecutive fillings, the reference measurements are performed with a finite but small residual gas pressure in the measurement cavity, $P_{Res}$. To take this into account in the evaluation procedure, it is assessed by the use of a pressure gauge. Hence, any drift in the residual pressure in the measurement cavity is corrected for by this. This implies that the linear and non-linear drifts of the assessed pressure in the measurement cavity in Eq. (35), i.e. the $\dot{P}_m^{(Res)}(t)$ and $\ddot{P}_m^{(Res)}(t)$, are expected to be insignificant with respect to the drifts of the pressure in the reference cavity. This implies that the terms that contain $\dot{P}_m^{(Res)}(t)$ and $\ddot{P}_m^{(Res)}(t)$ can be neglected with respect to those that represent the reference cavity and contain $\dot{P}_r^{(Res)}(t)$ and $\ddot{P}_r^{(Res)}(t)$.

## 4.3. Estimate of the conditions for drifts of the pressure in the reference cavity to provide errors that are equal to the benchmarks

Since the gas leakage (or outgassing) increases the pressure in the reference cavity monotonically, this will give rise to an error in the assessment of the change of the beat frequency. In addition, since the leakage into the reference cavity is exponential, the non-linear contributions are largest at the beginning of a measurement series. Under these conditions, both exponential functions in the Eqs. (49) and (50) can be considered to be close to unity. Hence, it can be concluded that $[\dot{P}_r^{(Res)}(t)/\beta]^2 \ll \ddot{P}_r^{(Res)}(t)/\beta$. This implies that under the condition that gas leakage (or outgassing) into the reference cavity is the only source of non-linear drift, the error in the assessment of the change of the beat frequency in the GAMOR methodology, $\delta[\Delta f(t_n,t_g,t_{n+1})]$ can be given by

$$\delta[\Delta f(t_n,t_g,t_{n+1})] = -\frac{1}{2}v_m^{(0)}\frac{\ddot{P}_r^{(Res)}}{\beta}(t_g-t_n)(t_{n+1}-t_g). \qquad (51)$$

By assuming, as above, that a "close-to" worst case condition takes place when $(t_0-t_n)(t_{n+1}-t_0) \approx (t_{n+1}-t_n)^2/5$ and $t \ll \tau_r$, this implies that $\delta[\Delta f(t_n,t_{n+1})]$ can be written as

$$\delta[\Delta f(t_n,t_{n+1})] \approx -\frac{(n-1)_{atm}}{10\tau_r^2}(t_{n+1}-t_n)^2 v_m^{(0)}, \qquad (52)$$

This implies, by the use of Eq. (7), that the error in refractivity from a leak (or outgassing) in the reference cavity, i.e. $\delta(n_m-1)$, can be expressed as

$$\delta(n_m-1) \approx \frac{\delta[\Delta f(t_n,t_{n+1})]}{v_m^{(0)}} \approx -\frac{(t_{n+1}-t_n)^2}{10\tau_r^2}(n-1)_{atm}. \qquad (53)$$

This shows that a leakage of gas into the reference cavity will contribute to an error in the refractivity that is equal to the strict benchmark condition whenever the leak has a characteristic time ($\tau_r$) that is $(t_{n+1}-t_n)10^4/\sqrt{10}$, which, for the case with a measurement cycle time of 100 s, corresponds to $3.5 \times 10^5$ s, or 4 days.

For the case with the relaxed benchmark condition, the corresponding characteristic leak time is one order of magnitude shorter, i.e. $3.5 \times 10^4$ s, or 10 hours.

These characteristic leak times correspond to maximum increase rates in pressure of 0.3 and 3 Pa/s, respectively. Since no well-constructed system should exhibit a leakage or outgassing rate of this magnitude, it is possible to conclude that it is unlikely that any gas leakage (or outgassing) in the reference cavity will influence any GAMOR measurement with errors close to any of the benchmarks.

The fact that the entity in Eq. (53) is a negative number shows that the error from a gas leak into the reference cavity, if it would appear, always would be negative.

Note that, for the case when the reference cavity is not evacuated but instead filled with gas of a given pressure, the leakage rates will be lower than modelled above. In those cases, the influence of gas leaks into the reference cavity can, as long as the cavity is filled with the same type of gas as is in the surrounding, be fully neglected.

## *4.4. Estimate of the conditions for leaks by other gases than that addressed by the GAMOR instrumentation*

Since different gases have dissimilar refractivity, a leakage of one gas (the surrounding gas) into the measurement cavity when the system run on another gas (e.g. $N_2$ or He) can be a source of error to the assessment. Since the refractivity of nitrogen and He are 1.7% and 87% lower than that of air, respectively, a refractometry assessment of $N_2$ or He in a surrounding of air will be affected by a gas leak even if the pressure provided is regulated by a pressure regulator (e.g. a piston balance).

In addition, due to their difference in refractivity, these types of assessment will be affected differently by a given leak. For the case with a measurement cycle of 100 s, to provide assessments below the strict benchmark condition, an assessment of refractivity of $N_2$ will require a leak rate into the measurement cavity that gives rise to an increase rate in pressure that is below 1 mPa/s. When refractometry is performed on He, the corresponding limiting pressure increase rate is significantly lower, in this case 0.02 mPa/s.

It is of importance to note that these requirements are significantly more strict than those for leakages (or outgassing) into the reference cavity by the same type of gas as addressed by the refractometer (as given in section 4.3 above). Hence, if not taken care of properly, this process can be the limiting source of error in GAMOR. One way of alleviating this is to place the vital parts of the refractometer in an enclosure containing the same gas as is addressed.

For the cases when unmodulated FP-based refractometry is performed, for which the time between the reference and the filled measurement cavity assessments often is orders of magnitude longer that what is used in GAMOR, the corresponding limiting leak rates are significantly lower. This emphasizes the importance of frequent gas exchange in FP-based refractometry systems.

## 5. Combined effect of drifts of the cavity lengths and gas leaks and outgassing

Equation (35) above also contains double products of the two effects considered above, i.e. products of the first order derivatives of the lengths of, and the pressures in, the two cavities. However, based on simple models of the drifts of the temperature of the cavity spacer and the residual gas in the reference cavity, in the same way as the terms containing the square of the first order derivatives of the cavity lengths or the pressure are inferior to the corresponding second order derivatives, the double products of the two effects considered above can be neglected with respect to the two second order derivatives.

## 6. Summary and conclusions

As has previously been shown [31, 32], the GAMOR methodology has an extraordinary ability to eliminate the influence of drifts in the system, including those originating from changes in the lengths of the cavities and those from leakage of gas and outgassing into the reference cavity. This work has provided a description of the principle behind the GAMOR methodology and explicates the background to this unique property.

To properly assess its properties and advantages, the conditions that limit the performance of GAMOR have been placed in perspective of those of two other types of FP-based types of refractometry: (conventional) single FP-cavity (SFPC) based refractometry and conventional (unmodulated) DFPC-based refractometry. It is shown that while SFPC-based refractometry is limited by the drifts in the measurement cavity, and conventional DFPC-based refractometry is restricted by the *difference* in drift of the two cavities, GAMOR is solely affected by the *difference in the non-linear parts* of the drifts of the two cavities!

This work has also provided an expression that can be used to estimate the contribution of the non-linear terms of various types of drift to the assessments of gas refractivity by GAMOR.

To assess to which degree drifts still can influence the GAMOR methodology, a benchmark for the errors (or uncertainties) in refractivity of $3 \times 10^{-12}$, corresponding to a pressure of $N_2$ of 1 mPa, (referred to as the strict benchmark) was defined.

Based on simple models of the temperature-induced drift of the lengths of the cavities and the residual gas in the reference cavity, this work then estimated the non-linear contribution of the drifts. It was estimated that when the GAMOR methodology is used for assessment of refractivity or density, it can sustain temperature fluctuations in the cavity spacer material of up to 100 mK if they have a characteristic time of $10^3$ s (estimated to be 30 times higher than what is required in conventional DFPC-based refractometry under similar conditions, 3 mK) before the errors and uncertainties exceed this benchmark.

However, for the case when the temperature fluctuations in the cavity spacer take place on faster time scales, the requirements are stronger. For example, for the case when the characteristic time for the temperature fluctuations is only $10^2$ s, the GAMOR methodology can be influenced with errors (or uncertainties) at the strict benchmark level if the temperature fluctuations are (still under worst conditions) solely 1 mK. Since this is non-trivial to achieve, this indicates that it is of importance to construct cavity spacer systems with a long characteristic time.

On the other hand, it can additionally be concluded that the GAMOR methodology is not influenced by the fastest changes in the length of the cavities, as for example takes place when the spacer material is exposed to sudden relaxations, since then only a single measurement point in an entire measurement series (corresponding to a single measurement cycle) will be affected, giving rise to a spurious outlier. Such an outlier can be eliminated by, for example, applying a median filter to a series of consecutive assessments.

The GAMOR methodology can also accommodate a reference cavity with a characteristic leak rate of 4 days (which corresponds to an increase in pressure of 0.3 Pa/s) before the measurement error, under a set of worst conditions, exceeds the same benchmark condition.

Since well-designed systems have temperature fluctuations and leakage rates that are smaller than these, it is concluded that, as long as the temperature fluctuations have a characteristic time of at least $10^3$ s and the gas around the refractometer is the same at that addressed (have the same index of refraction), there will not be any appreciable influence of non-linear drifts to refractivity (or density) assessments made by the GAMOR methodology. Hence, it was concluded that such a GAMOR system can sustain significant temperature drifts and considerable leakage or outgassing rates without being affected by noticeable errors (or uncertainties) when refractivity and density are assessed.

It is prophesized though that the most severe drifts that GAMOR can be affected by are those from leaks of environmental gas in the system when other gases are addressed. For the case with a measurement cycle of 100 s, an assessment of refractivity of $N_2$ in an environment of air with an error below the strict benchmark will require a leak rate into the measurement cavity that gives rise to an increase rate in pressure that is below 1 mPa/s. For He, the corresponding limiting increase rate in pressure is 0.02 mPa/s. Unless taken care of properly, this can potentially be the limiting source of error in GAMOR. One way to alleviate this is to place the vital parts of the refractometer in an enclosure containing the same gas as is addressed.

It is worth no note though that the methodology cannot eliminate the error (or uncertainty) in pressure assessments that originates from undetected temperature fluctuations around the measurement cavity, or the uncertainty of the temperature assessment equipment, through the equation of state. Such fluctuations/uncertainties, $\delta(T_m)$, provide a contribution to the relative error, i.e. $\delta(P_m)/P_m$, that is given by $\delta(T_m)/T_m$. This implies that, for high pressures (predominantly those in the kPa range), the errors (or uncertainties) in the assessment of pressure from drifts of the temperature related to the equation of state will dominate over those from drifts of the cavity lengths.

In conclusion, the analysis presented here shows that the GAMOR methodology in practice is capable of eliminating most types of drifts that can appear in FP-cavity-based refractometry systems. This opens up for the fact that GAMOR systems, when properly calibrated, can be used for highly accurate assessments of refractivity, gas density, and pressure under a variety of conditions. It also illustrates that, in comparison with ordinary FP-refractometry, instrumentations based on the GAMOR methodology can, if needed, be realized around experimental set-ups with relaxed conditions regarding drifts of temperature, material creeping, gas leaks, and outgassing, allowing for transportable instrumentation [34]. All this demonstrates clearly the advantages of realizing GAMOR in FP-based refractometry instrumentation.

**Acknowledgements**

This research was in part supported by the Swedish Research Council (VR), project number 621-2015-04374; the Umeå University Industrial doctoral school (IDS); the Vinnova Metrology Programme, project numbers 2017-05013 and 2018-04570; and the Kempe Foundations, project number 1823, U12. We are also grateful to Thomas Hausmaninger for fruitful discussions.

**References**

[1] Andersson, M., Eliasson, L., and Pendrill, L.R. 1987 Compressible Fabry-Perot refractometer, *Appl. Opt.* **26** 4835-4840 https://doi.org/10.1364/AO.26.004835

[2] Khelifa, N., Fang, H., Xu, J., Juncar, P., and Himbert, M. 1998 Refractometer for tracking changes in the refractive index of air near 780 nm, *Appl. Opt.* **37** 156-161 https://doi.org/10.1364/AO.37.000156

[3] Axner, O., Silander, I., Hausmaninger, T., and Zelan, M. 2017 Drift-free Fabry-Perot-cavity-based optical refractometry – accurate expressions for assessments of refractivity and gas density from the change in laser frequency that follows a cavity evacuation, arXiv:1704.01187
https://arxiv.org/abs/1704.01187v2

[4] Pendrill, L.R. 1988 Density of moist air monitored by laser refractometry, *Metrologia* **25** 87-93
https://doi.org/10.1088/0026-1394/25/2/005

[5] Fang, H. and Juncar, P. 1999 A new simple compact refractometer applied to measurements of air density fluctuations, *Rev. Sci. Instrum.* **70** 3160-3166
https://doi.org/10.1063/1.1149880

[6] Fang, H., Picard, A., and Juncar, P. 2002 A heterodyne refractometer for air index of refraction and air density measurements, *Rev. Sci. Instrum.* **73** 1934-1938 https://doi.org/10.1063/1.1459091

[7] Pendrill, L.R. 2004 Refractometry and gas density, *Metrologia* **41** S40-S51 https://doi.org/10.1088/0026-1394/41/2/s04

[8] Hedlund, E. and Pendrill, L.R. 2006 Improved determination of the gas flow rate for UHV and leak metrology with laser refractometry, *Meas. Sci. Technol.* **17** 2767-2772 https://doi.org/10.1088/0957-0233/17/10/031

[9] Hedlund, E. and Pendrill, L.R. 2007 Addendum to "Improved determination of the gas flow rate for UHV and leak metrology with laser refractometry (vol 17, pg 2767, 2006)", *Meas. Sci. Technol.* **18** 3661-3663 https://doi.org/10.1088/0957-0233/18/11/052

[10] Egan, P. and Stone, J.A. 2011 Absolute refractometry of dry gas to +/- 3 parts in $10^9$, *Appl. Opt.* **50** 3076-3086 https://doi.org/10.1364/ao.50.003076

[11] Silander, I., Zelan, M., Axner, O., Arrhen, F., Pendrill, L., and Foltynowicz, A. 2013 Optical measurement of the gas number density in a Fabry-Perot cavity, *Meas. Sci. Technol.* **24** 105207 https://doi.org/10.1088/0957-0233/24/10/105207

[12] Mari, D., Bergoglio, M., Pisani, M., and Zucco, M. 2014 Dynamic vacuum measurement by an optical interferometric technique, *Meas. Sci. Technol.* **25** https://doi.org/10.1088/0957-0233/25/12/125303

[13] Yan, L.P., Chen, B.Y., Zhang, E.Z., Zhang, S.H., and Yang, Y. 2015 Precision measurement of refractive index of air based on laser synthetic wavelength interferometry with Edlen equation estimation, *Rev. Sci. Instrum.* **86** 085111 https://doi.org/10.1063/1.4928159

[14] Egan, P.F., Stone, J.A., Ricker, J.E., and Hendricks, J.H. 2016 Comparison measurements of low-pressure between a laser refractometer and ultrasonic manometer, *Rev. Sci. Instrum.* **87** 053113 https://doi.org/10.1063/1.4949504

[15] Silander, I., Hausmaninger, T., Zelan, M., and Axner, O. 2017 Fast switching dual Fabry-Perot cavity optical refractometry - methodologies for accurate assessment of gas density, arXiv:1704.01186v01182 https://arxiv.org/abs/1704.01186v2

[16] Zelan, M., Silander, I., Hausmaninger, T., and Axner, O. 2017 Drift-Free Fabry-Perot-Cavity-based Optical Refractometry - Estimates of its Precision, Accuracy, and Temperature dependence, arXiv:1704.01185v01182 https://arxiv.org/abs/1704.01185v2

[17] Hendricks, J.H., Strouse, G.F., Ricker, J.E., Olson, D.A., Scace, G.E., Stone, J.A., and Egan, P.F. 2015 Photonic article, process for making and using same, **US20160018280 A1**, https://patents.google.com/patent/US20160018280

[18] Matthews, J.C.F., Zhou, X.Q., Cable, H., Shadbolt, P.J., Saunders, D.J., Durkin, G.A., Pryde, G.J., and O'Brien, J.L. 2016 Towards practical quantum metrology with photon counting, *Npj Quantum Information* **2** https://doi.org/10.1038/npjqi.2016.23

[19] Egan, P.F., Stone, J.A., Ricker, J.E., Hendricks, J.H., and Strouse, G.F. 2017 Cell-based refractometer for pascal realization, *Opt. Lett.* **42** 2944-2947 https://doi.org/10.1364/ol.42.002944

[20] Jousten, K., Hendricks, J., Barker, D., Douglas, K., Eckel, S., Egan, P., Fedchak, J., Flugge, J., Gaiser, C., Olson, D., Ricker, J., Rubin, T., Sabuga, W., Scherschligt, J., Schodel, R., Sterr, U., Stone, J., and Strouse, G. 2017 Perspectives for a new realization of the pascal by optical methods, *Metrologia* **54** S146-S161 https://doi.org/10.1088/1681-7575/aa8a4d

[21] Silvestri, Z., Boineau, F., Otal, P., and Wallerand, J.P., 'Helium-based refractometry for pressure measurements in the range 1-100 kPa', in IEEE (ed.), *2018 Conference on Precision Electromagnetic Measurements*, (IEEE, New York, NY 10017, USA, 2018) https://doi.org/10.1109/CPEM.2018.8501259

[22] Scherschligt, J., Fedchak, J.A., Ahmed, Z., Barker, D.S., Douglass, K., Eckel, S., Hanson, E., Hendricks, J., Klimov, N., Purdy, T., Ricker, J., Singh, R., and Stone, J. 2018 Review Article: Quantum-based vacuum metrology at the National Institute of Standards and Technology, *Journal of Vacuum Science & Technology A* **36** https://doi.org/10.1116/1.5033568

[23] Mari, D., Pisani, M., and Zucco, M. 2019 Towards the realization of an optical pressure standard, *Measurement* **132** 402-407 https://doi.org/10.1016/j.measurement.2018.09.069

[24] Eickhoff, M.L. and Hall, J.L. 1997 Real-time precision refractometry: New approaches, *Appl. Opt.* **36** 1223-1234 https://doi.org/10.1364/ao.36.001223

[25] Fox, R.W., Washburn, B.R., Newbury, N.R., and Hollberg, L. 2005 Wavelength references for interferometry in air, *Appl. Opt.* **44** 7793-7801 https://doi.org/10.1364/ao.44.007793

[26] Xiao, G.Z., Adnet, A., Zhang, Z.Y., Sun, F.G., and Grover, C.P. 2005 Monitoring changes in the refractive index of gases by means of a fiber optic Fabry-Perot interferometer sensor, *Sensors and Actuators a-Physical* **118** 177-182 https://doi.org/10.1016/j.sna.2004.08.029

[27] Stone, J.A. and Stejskal, A. 2004 Using helium as a standard of refractive index: correcting errors in a gas refractometer, *Metrologia* **41** 189-197 https://doi.org/10.1088/0026-1394/41/3/012

[28] Fox, R.W. 2009 Temperature analysis of low-expansion Fabry-Perot cavities, *Opt. Express* **17** 15023-15031 https://doi.org/10.1364/oe.17.015023

[29] Egan, P.F., Stone, J.A., Hendricks, J.H., Ricker, J.E., Scace, G.E., and Strouse, G.F. 2015 Performance of a dual Fabry-Perot cavity refractometer, *Opt. Lett.* **40** 3945-3948 https://doi.org/10.1364/ol.40.003945

[30] Rubin, T. and Yang, Y. 2018 Simulation of pressure induced length change of an optical cavity used for optical pressure standard, *Journal of Physics: Conf. Series* **1065** 162003 https://doi.org/10.1088/1742-6596/1065/16/162003

[31] Silander, I., Hausmaninger, T., Zelan, M., and Axner, O. 2018 Gas modulation refractometry for high-precision assessment of pressure under non-temperature-stabilized conditions, *Journal of Vacuum Science and Technology A* **36** 03E105 https://doi.org/10.1116/1.5022244

[32] Silander, I., Hausmaninger, T., Forssén, C., Zelan, M., and Axner, O. 2019 Gas Eqilibration gas modulation refractometry (GEq-GAMOR) for assessment of pressure with sub-ppm precision, *Submitted for publication to JVST A*

[33] Avdiaj, S., Yang, Y.C., Jousten, K., and Rubin, T. 2018 Note: Diffusion constant and solubility of helium in ULE glass at 23 degrees C, *J. Chem. Phys.* **148** 116101 https://doi.org/10.1063/1.5019015

[34] Forssén, C., Silander, I., Hausmaninger, T., Axner, O., and Zelan, M. 2019 Transportable GAMOR system, *In preparation*